\documentclass[preprint,12pt]{elsarticle}

\usepackage{amssymb}
\usepackage{mathrsfs}
\usepackage{amsmath}

\begin{document}
\begin{frontmatter}

\title{Stability band structure for periodic states in periodic potentials}
\author{Bin Liu$^{1}$ }
\author{Lu Li$^{1,*}$}
\ead{llz@sxu.edu.cn}
\author{Boris A. Malomed$^{2,3}$}
\address{$^{1}$Institute of Theoretical Physics, Shanxi University, Taiyuan 030006, China}
\address{$^{2}$Department of Physical Electronics, School of Electrical Engineering,
Faculty of Engineering, Tel Aviv University, Tel Aviv 69978, Israel}
\address{$^{3}$ITMO University, St. Petersburg 197101, Russia}

\begin{abstract}
A class of periodic solutions of the nonlinear Schr\"{o}dinger equation with
non-Hermitian potentials are considered. The system may be implemented in
planar nonlinear optical waveguides carrying an appropriate distribution of
local gain and loss, in a combination with a photonic-crystal structure. The
complex potential is built as a solution of the inverse problem, which
predicts the potential supporting required periodic solutions. The main
subject of the analysis is the spectral structure of the linear (in)stability
for the stationary spatially periodic states in the periodic potentials. The
stability and instability bands are calculated by means of the
plane-wave-expansion method, and verified in direct simulations of the
perturbed evolution. The results show that the periodic solutions may be
stable against perturbations in specific Floquet-Bloch bands, even if they are
unstable against small random perturbations.
\end{abstract}

\end{frontmatter}
\section{Introduction}

Non-Hermitian (complex) potentials in wave equations give rise to effects that
cannot be realized with Hermitian (real) potentials, well-known examples being
the parity-time ($\mathcal{PT}$) symmetry in the case when the real and
imaginary parts of the complex potential are, respectively, spatially even and
odd \cite{PRL80_5243,JMP43_2814,PRL89_270401,Rep.Prog.Phys70_947}. The
$\mathcal{PT}$ symmetry, i.e., the reality of the energy spectrum in the
respective systems, holds up to a critical value of the strength of the
imaginary potential, \emph{symmetry breaking }occurring above the critical
value. In optics, $\mathcal{PT}$-symmetric potentials have been realized in
various experiments
\cite{Guo,NP6_192,Nat488_167,NatMater12_108,Microres,NP10_394,atomic
chain,Optica}. They exhibit remarkable properties and potential applications,
such as power oscillations \cite{NP6_192}, non-reciprocal light propagation
\cite{PRL100_103904}, optical transparency \cite{PRL103_093902}, negative
refraction \cite{PRL113_023902}, pseudo-Hermitian Bloch oscillations
\cite{PRL103_123601,SR5_17760}, unidirectional invisibility
\cite{PRA82_043803,PRL106_213901,PRL110_234101}, and possibilities to design
various $\mathcal{PT}$-symmetric devices
\cite{PRL112_143903,Science346_972,Science346_975,PRL117_224302}, as recently
reviewed in Ref. \cite{NatPhys}. Further, optics provides a fertile ground to
investigate $\mathcal{PT}$-symmetric beam dynamics in nonlinear regimes,
including the formation of bright and dark solitons, gap solitons, defect
states, multi-peak modes, and vortices, see original works
\cite{18,19,Mirosh,22,20,21,32,23,24,OL37_4543,25,26,PRA87_013812,27,29,30,28,31,Raymond,OC313_139,OC315_303,PRA90_034833,OC335_146}
and recent reviews \cite{RMP,Suchkov}. In particular, a specific arrangement
of the self-defocusing nonlinearity, growing from the center to periphery,
makes it possible to predict self-trapped modes featuring \emph{unbreakable}
$\mathcal{PT}$ symmetry \cite{unbreakable}. The concept of the $\mathcal{PT}%
$-symmetry has also been applied to Bose-Einstein condensates
\cite{PRA86_013612,PRA90_042123,PRA93_033617,PRA91_043629}, atomic cells
\cite{OL38_4033,OL39_5387,OE21_32053}, and nonlinearity-induced $\mathcal{PT}%
$-symmetry without material gain \cite{NJP18_065001}.

Stability of optical solitons and nonlinear beam dynamics in non-$\mathcal{PT}%
$-symmetric complex potentials was also addressed, showing that the solitons
may be stable in a wide range of parameters
\cite{OL41_2747,SH,PhysicaD331_48,PLA380_3803,RMP}. Some applications, such as
coherent perfect absorbers and time-reversal lasers, have been elaborated in
such settings
\cite{PRL105_053901,Science331_889,PRL108_173901,NC5_4034,science346_328}, and
non-$\mathcal{PT}$-symmetric optical potentials with all-real spectra in a
coherent atomic system have been realized \cite{PRA95_023833}. The studies of
the stability of modes supported by the complex potentials make this topic a
part of the very broad field of dynamical stability in various nonlinear
systems. One of basic problems in this field is the modulation instability
(MI) of extended states.

In particular, the MI was widely studied in $\mathcal{PT}$-symmetric nonlinear
Schr\"{o}dinger (NLS) equations
\cite{PRE83_036608,OL36_4323,J.Opt15_064010,oe22_19774,PRE91_023203,PhysicaD313_26,PRE92_022913}%
. Recently, the MI of constant-amplitude waves has been addressed in models
with more general complex potentials by using the plane-wave-expansion method
combined with direct simulations \cite{NC6_7257,EPJD71_140}, which makes it
possible to calculate the stability band structure of spatially periodic
solutions in periodic complex potentials. In the present work, we aim to
explore this structure for periodic solutions of the NLS equation with
non-Hermitian potentials.

The paper is organized as follows. In the next section, the model and its
reduction are introduced, and the corresponding periodic solutions are
presented by solving an inverse problem, which predicts the periodic
potentials supporting a required phase-gradient structure of the periodic
solutions. In Sec. III, we focus on the analysis of the stability band
structure of the periodic solutions, employing the plane-wave-expansion
method. The conclusions are made in Sec. IV.
\section{Model and periodic solutions}

We begin the analysis by considering the NLS equation with non-Hermitian
potentials, written in a scaled form, cf. Ref. \cite{NC6_7257}:
\begin{equation}
i\frac{\partial\Psi}{\partial z}+\frac{\partial^{2}\Psi}{\partial x^{2}%
}+V(x)\Psi+g\left\vert \Psi\right\vert ^{2}\Psi=0. \label{Model}%
\end{equation}
In the application to light propagation in planar waveguides, $\Psi(x,z)$ is
the slowly varying envelope of the electric field, $z$ and $x$ are the
propagation distance and the transverse coordinate,
\begin{equation}
V(x)\equiv V_{R}(x)+iV_{I}(x) \label{V}%
\end{equation}
is the complex potential, which can be implemented in optics by combining the
spatially modulated refractive index and spatially distributed gain and loss
elements \cite{NP6_192}. The nonlinearity can be either self-focusing ($g>0$)
or defocusing ($g<0)$, the latter being possible in semiconductor optical
materials \cite{semicond}.
We are looking for stationary solutions of Eq. (\ref{Model}) as
\begin{equation}
\Psi(x,z)=\Phi(x)\exp(i\mu z), \label{Solution1}%
\end{equation}
where $\mu$ is a real propagation constant and complex field profile $\Phi(x)$
is determined by the following nonlinear equation:%
\begin{equation}
-\mu\Phi+\Phi^{^{\prime\prime}}+V(x)\Phi+g\left\vert \Phi\right\vert ^{2}%
\Phi=0, \label{Model2}%
\end{equation}
with the prime standing for $d/dx$. Further, we define real amplitude and
phase%
\begin{equation}
\Phi(x)=H\left(  x\right)  \exp\left[  i\Theta(x)\right]  , \label{Solution2}%
\end{equation}
for which complex Eq. (\ref{Model2}) splits into real ones:%
\begin{align}
-\mu H+H^{^{\prime\prime}}-H(\Theta^{^{\prime}})^{2}+V_{R}H+gH^{3}  &
=0,\label{Real part}\\
2H^{^{\prime}}\Theta^{\prime}+H\Theta^{^{\prime\prime}}+V_{I}H  &  =0.
\label{Imag part}%
\end{align}
Equations (\ref{Real part}) and (\ref{Imag part}) may be addressed as an
inverse problem, which provides a required form of the solution, $H(x)$ and
$\Theta(x)$, by selecting a particular complex potential (\ref{V}):
\begin{align}
V_{R}\left(  x\right)   &  =\mu-\frac{H^{^{\prime\prime}}}{H}+(\Theta
^{^{\prime}})^{2}-gH^{2},\label{vr1}\\
V_{I}\left(  x\right)   &  =-2\frac{H^{^{\prime}}\Theta^{\prime}}{H}%
-\Theta^{^{\prime\prime}}. \label{vi1}%
\end{align}
The approach based on the inverse problem was previously elaborated in various
contexts related to NLS equations
\cite{inverse1,inverse2,inverse3,inverse4,inverse5}.

To choose basic periodic solutions to Eq. (\ref{Model}), avoiding
singularities in the complex potential, we set $H^{^{\prime\prime}}%
/H=-\omega^{2}$ and $\Theta^{\prime}=V_{0}H$, where $\omega$ and $V_{0}$ are
the real constants. Accordingly, we have
\begin{align}
H(x)  &  =A\sin(\omega x+\phi),\label{H}\\
\Theta(x)  &  =-\frac{V_{0}A}{\omega}\cos\left(  \omega x+\phi\right)  ,
\label{Theta}%
\end{align}
with arbitrary real constants $A$ and $\phi$. Thus, the starting point is the
periodic solution of Eq. (\ref{Model}) taken as
\begin{gather}
\Psi(x,z)=H\left(  x\right)  e^{i\Xi(x,z)},\label{Solution}\\
\Xi(x,z)\equiv\mu z-\frac{AV_{0}}{\omega}\cos(\omega x+\phi),\nonumber
\end{gather}
with the corresponding real and imaginary parts of the complex potential given
by Eqs. (\ref{vr1}) and (\ref{vi1}):
\begin{align}
V_{R}\left(  x\right)   &  =\mu+\omega^{2}+\left(  V_{0}^{2}-g\right)
H^{2}\left(  x\right)  ,\label{VR}\\
V_{I}\left(  x\right)   &  =-3V_{0}H^{\prime}\left(  x\right)  . \label{VI}%
\end{align}

The complex periodic potential $V\left(  x\right)  $ given by Eqs. (\ref{VR})
and (\ref{VI}) is similar to the so-called Wadati potential \cite{VVK}, with
even and odd real and imaginary parts, respectively. Note that the periodic
solution exists in the linear limit ($g=0$), as well as for an arbitrary
strength of the nonlinearity ($g\neq0$). Lastly, function $H\left(  x\right)
$ determines the power flow from gain to loss regions, the respective Poynting
vector, $S=(i/2)(\Psi\partial\Psi^{\ast}/\partial x-\Psi^{\ast}\partial
\Psi/\partial x)$, taking a very simple form, $S=V_{0}H^{3}$, proportional to
the gain-loss strength, $V_{0}$.

\begin{figure}[ptb]
\centering\vspace{-0.5cm} \includegraphics[width=9.0cm]{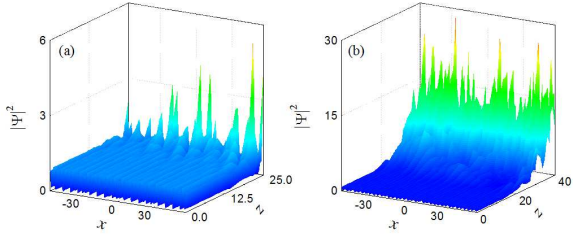}
\vspace{-3.5cm}\caption{(Color online) The evolution of the perturbed periodic
solution under the action of the periodic complex potential $V(x)$ given by
Eqs. (\ref{VR}) and (\ref{VI}) for (a) the self-focusing nonlinearity with
$g=0.3$ and $V_{0}=0.23$; and (b) the defoucsing nonlinearity with $g=-0.7$
and $V_{0}=0.2$. Here, the perturbation amplitude $\varepsilon=0.02$, and the
other parameters are taken as $A=1$, $\omega=1$, $\phi=\pi/2$, and $\mu=-1$.}%
\end{figure}

\section{The stability-band structure of the periodic solution}

In this section, we address the stability of the periodic solution
(\ref{Solution}), using the plane-wave-expansion method for the
linear-stability analysis. The results will be verified by means of direct
numerical simulations.

The linear-stability analysis is initiated by adding a small perturbation to
periodic solution (\ref{Solution}):
\begin{equation}
\Psi(x,z)=\left[  H+\varepsilon F_{\lambda}(x)e^{i\lambda z}+\varepsilon
G_{\lambda}^{\ast}(x)e^{-i\lambda^{\ast}z}\right]  e^{i\Xi(x,z)}, \label{PP}%
\end{equation}
where $\ast$ stands for the complex conjugate, and $\varepsilon$ is a real
infinitesimal amplitude of the perturbation with complex eigenfunctions
$F_{\lambda}\left(  x\right)  $ and $G_{\lambda}\left(  x\right)  $, which are
related to complex eigenvalue $\lambda$. As usual, an imaginary part of
$\lambda$, if any, defines the instability growth rate of the perturbation.
The substitution of expression (\ref{PP}) into Eq. (\ref{Model}) and
subsequent linearization leads to the eigenvalue problem in the matrix form,
\begin{equation}
\left(
\begin{array}
[c]{cc}%
L_{1} & gH^{2}\\
-gH^{2} & L_{2}%
\end{array}
\right)  \left(
\begin{array}
[c]{c}%
F_{\lambda}\left(  x\right) \\
G_{\lambda}\left(  x\right)
\end{array}
\right)  =\lambda\left(
\begin{array}
[c]{c}%
F_{\lambda}\left(  x\right) \\
G_{\lambda}\left(  x\right)
\end{array}
\right)  , \label{Eigenfunction}%
\end{equation}
where the operators $L_{1}$ and $L_{2}$ are
\begin{align}
L_{1}  &  =-\mu+gH^{2}-2iV_{0}H_{x}+2iV_{0}H\partial_{x}+\partial
_{xx},\label{L1}\\
L_{2}  &  =\mu-gH^{2}-2iV_{0}H_{x}+2iV_{0}H\partial_{x}-\partial_{xx}.
\label{L2}%
\end{align}

The linear eigenvalue problem (\ref{Eigenfunction}) can be solved by the
finite difference method. As typical examples, we calculated the eigenvalues
of Eq. (\ref{Eigenfunction}), finding that the instability growth rates are
$0.0036$ and $0.5673$ for parameters given in the caption to Fig. 1, for the
self-focusing and defocusing nonlinearity, respectively, which means that the
periodic solution (\ref{Solution}) is unstable. To confirm the results, we
have performed simulations of the evolution of solution (\ref{Solution}) with
random-noise perturbations added to it, as shown in Figs. 1(a) and 1(b).

The above results do not include the stability band structure of the periodic
solution (\ref{Solution}). Below, we will apply the plane-wave-expansion
method based on Eq. (\ref{Eigenfunction}) \cite{NC6_7257} to produce the band
structure. In the framework of this method in its general form, because $H(x)$
is a periodic function with period of $2\pi/\omega$, the perturbation
eigenmodes $F_{\lambda}\left(  x\right)  $ and $G_{\lambda}\left(  x\right)
$, along with $H(x)$ itself, are expanded into Fourier series, according to
the Floquet-Bloch theorem:
\begin{align}
\left(
\begin{array}
[c]{c}%
F_{\lambda}\left(  x\right) \\
G_{\lambda}\left(  x\right)
\end{array}
\right)   &  =%
%TCIMACRO{\dsum _{n=-\infty}^{+\infty}}%
%BeginExpansion
{\displaystyle\sum_{n=-\infty}^{+\infty}}
%EndExpansion
\left(
\begin{array}
[c]{c}%
u_{n}(k)\\
v_{n}(k)
\end{array}
\right)  e^{i(n\omega+k)x},\label{FG_pw}\\
H(x)  &  =%
%TCIMACRO{\dsum _{n=-\infty}^{+\infty}}%
%BeginExpansion
{\displaystyle\sum_{n=-\infty}^{+\infty}}
%EndExpansion
H_{n}e^{in\omega x}, \label{W_pw}%
\end{align}
where $k$ is the Bloch momentum, making the eigenmodes quasiperiodic functions
of $x$. Substituting Eqs. (\ref{VR}), (\ref{VI}), (\ref{FG_pw}) and
(\ref{W_pw}) into the eigenvalue problem (\ref{Eigenfunction}), one arrives at
the following system of linear equations for perturbation coefficients $u_{n}%
$, $v_{n}$ and eigenvalue $\lambda(k)$:%
\begin{align}
\lambda(k)u_{n}  &  =\delta u_{n-2}+\delta v_{n-2}+\beta_{n-1}u_{n-1}%
\nonumber\\
&  +\alpha_{n}u_{n}+\gamma_{n+1}u_{n+1}+\sigma u_{n+2}+\sigma v_{n+2}%
,\nonumber\\
\lambda(k)v_{n}  &  =-\delta u_{n-2}-\delta v_{n-2}+\beta_{n-1}v_{n-1}%
\nonumber\\
&  -\alpha_{n}v_{n}+\gamma_{n+1}v_{n+1}-\sigma u_{n+2}-\sigma v_{n+2},
\label{PT_UV}%
\end{align}
where we define
\begin{align}
\delta &  =-\frac{1}{4}gA^{2}e^{i2\phi},\\
\sigma &  =-\frac{1}{4}gA^{2}e^{-i2\phi},\\
\alpha_{n}  &  =-\mu-(n\omega+k)^{2},\\
\beta_{n}  &  =-iAV_{0}[\left(  n+1\right)  \omega+k]e^{-i\phi},\\
\gamma_{n}  &  =iAV_{0}[\left(  n-1\right)  \omega+k]e^{i\phi}.
\end{align}
The instability growth rate of the periodic solution is again defined as the
largest imaginary part of $\lambda(k)$, in the set of the eigenvalues for
given $k$. As a typical example, Fig. 2 depicts the dependence of
$\max[\operatorname{Im}\lambda(k)]$ on $k$ (in the half of the first Brillouin
zone) for different values of the gain/loss parameter $V_{0}$ in both the
self-focusing and defocusing regimes.

\begin{figure}[ptb]
\centering\vspace{0.0cm} \includegraphics[width=8.5cm]{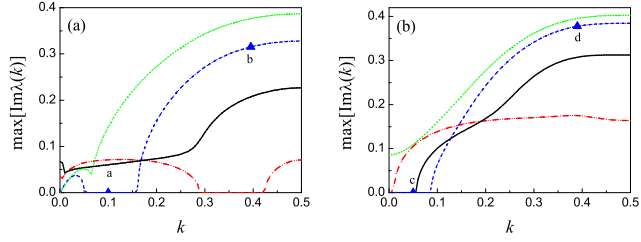}
\vspace{-0.0cm}\caption{(Color online) The dependence of the instability
growth rate, i.e., the largest imaginary part of eigenvalues $\lambda(k)$, on
the Bloch wavenumber $k$ (in the half of the first Brillouin zone) for
different values of the gain/loss coefficient $V_{0}$. Panels (a) and (b)
report the results of the self-focusing ($g=0.3$) and defocusing ($g=-0.7$)
nonlinearity, respectively. Red dashed, black solid, blue short-dashed, and
green short-dotted curves correspond to $V_{0}=$ $0.05$, $0.16$, $0.23$, and
$0.27$ in (a), and $V_{0}=0.08$, $0.16$, $0.20$, and $0.21$ in (b),
respectively. Other parameters are the same as in Fig. 1.}%
\end{figure}

For the self-focusing nonlinearity, as seen in Fig. 2(a), at $V_{0}=0.16$ and
$V_{0}=0.27$ the eigenvalues with the largest imaginary part are complex at
all $k$ [see the black solid and green short-dotted curves in Fig. 2(a)],
hence the periodic solution is linearly unstable to all perturbations. When
$V_{0}=0.05$ and $0.23$, as shown by the red dashed and blue short-dashed
curves in Fig. 2(a), a stability band, i.e., an interval of wavenumber $k$ in
which the instability growth rate vanishes, can be formed. This means that the
periodic solution is stable against perturbations corresponding to the
Floquet-Bloch modes in intervals $k\in\lbrack0.29,0.42]$ as $V_{0}=0.05$, and
$[0.055,0.155]$ as $V_{0}=0.23$, respectively.

Similarly, for the defocusing nonlinearity, as shown in Fig. 2(b), at
$V_{0}=0.08$ and $V_{0}=0.21$, the eigenvalues with the largest imaginary part
are complex at all $k$ [see the red dashed and green short dotted curves in
Fig. 2(b)]. On the other hand, at $V_{0}=0.16$ and $0.2$ stability bands are
$[0,0.055]$ and $[0,0.085]$, respectively. They are narrower than the
instability zone, and are mainly located near the center of the Brillouin zone
[see the black solid and blue short-dashed curves in Fig. 2(b)].

\begin{figure}[ptb]
\centering\vspace{0.0cm} \includegraphics[width=8.5cm]{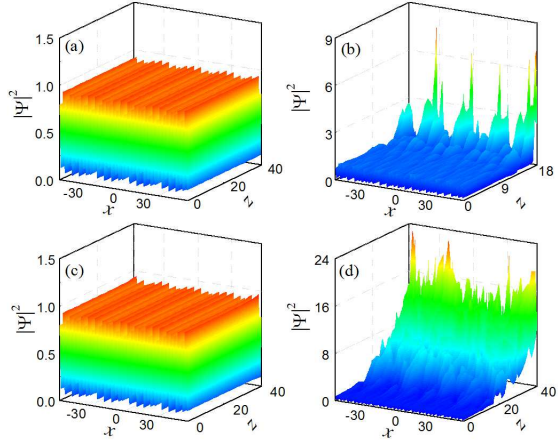}
\vspace{-0.0cm}\caption{(Color online) The simulated evolution of the periodic
solution, perturbed by specific eigenmodes (see the text), under the action of
the self-focusing nonlinearity ($g=0.3$) and periodic potential ($V_{0}=0.23$)
in (a) and (b), and defocusing nonlinearity ($g=-0.7$) and periodic potential
($V_{0}=0.2$) in (c) and (d). Here, (a) $k=0.1$; (b) $k=0.39$; (c) $k=0.05$,
and (d) $k=0.39$. The perturbation amplitude is $\varepsilon=0.02$, and other
parameters are the same as in Fig. 1.}%
\end{figure}

To verify the above results, we have performed systematic simulations of Eq.
(\ref{Model}) by taking inputs in the form of the periodic solution with the
addition of small perturbations corresponding to specific Floquet-Bloch
eigenmodes, as per Eq. (\ref{PP}). Figs. 3(a) and 3(b) show the evolution of
the periodic solution perturbed by the eigenmodes corresponding to $k=0.1$ and
$k=0.39$ at $V_{0}=0.23$ in the self-focusing regime. As predicted by points
\textquotedblleft a\textquotedblright\ and \textquotedblleft
b\textquotedblright\ in Fig. 2(a), the periodic solution is stable against
these perturbation modes at $k=0.1$ [Fig. 3(a)], and unstable at $k=0.39$
[Fig. 3(b)]. Figs. 3(c) and 3(d) display the evolution in the system with
$V_{0}=0.2$ and the defocusing nonlinearity, where the perturbation eigenmodes
corresponding to $k=0.05$ and $k=0.39$ are initially added, respectively. They
demonstrate that, in agreement with the prediction of points \textquotedblleft
c\textquotedblright\ and \textquotedblleft d\textquotedblright\ in Fig. 2(b),
the periodic solution is stable to the perturbations at $k=0.05$, see Fig.
3(c), and it is unstable to $k=0.39$, as shown in Fig. 3(d).

\begin{figure}[ptb]
\centering\vspace{0.0cm} \includegraphics[width=8.5cm]{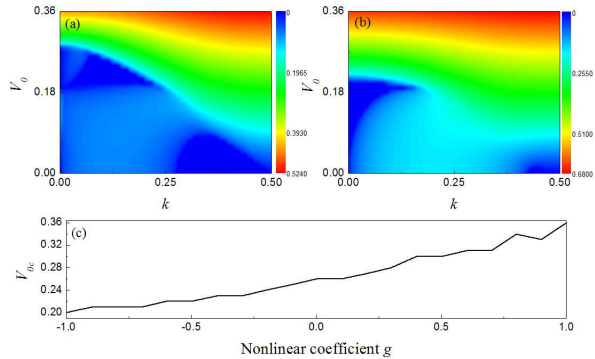}
\vspace{-0.0cm}\caption{(Color online) The dependence of the largest imaginary
part of the eigenvalues on $V_{0}$ in the half of first Brillouin zone for (a)
$g=0.3$ and (b) $g=-0.7$, respectively. (c) The critical value $V_{0c}$,
beyond which the stability band does not exist at all $k$, versus the
nonlinearity coefficient $g$. Other parameters are the same as in Fig. 1.}%
\end{figure}

To exhibit the influence of the gain/loss parameter $V_{0}$ on the stability
band structure, Fig. 4 presents the largest imaginary part of the eigenvalues
vs. $V_{0}$ in the half of first Brillouin zone for both the self-focusing and
defocusing nonlinearities. For the self-focusing case, as shown in Fig. 4(a),
the stability band does not exist at all $k$ when $V_{0}$ belongs to intervals
$0.08<V_{0}<0.2$ and $0.26<V_{0}<0.36$, while in intervals $0\leq V_{0}%
\leq0.08$ and $0.2\leq V_{0}\leq0.26$ the stability band appears in some
interval of $k$, which shrinks with the increase of $V_{0}$. Fig. 4(b) shows
the defocusing case. Similarly, we find that the stability band does not exist
at all $k$ when $V_{0}\in(0.02,0.08)\cup(0.20,0.36)$, while in $[0.08,0.20]$
the stability band appears in some interval of $k$, chiefly near the center of
the Brillouin zone. For the self-focusing or defocusing nonlinearity alike,
there exists a critical value $V_{0c}$ such that, at $V_{0}>V_{0c}$, stability
banks do not exist. Fig. 4(c) presents the dependence of critical value
$V_{0c}$ on the nonlinearity coefficient, $g$. Naturally, $V_{0c}$ increases
with $g$, as the interplay of the nonlinearity with the $\mathcal{PT}$
symmetry usually accelerates the onset of the breakup of the symmetry.

\begin{figure}[ptb]
\centering\vspace{0.0cm} \includegraphics[width=8.5cm]{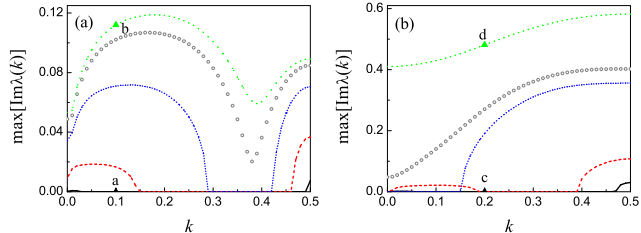}
\vspace{-0.0cm}\caption{(Color online) The dependence of the instability
growth rate, $\max[\operatorname{Im}\lambda(k)]$, on Bloch wavenumber $k$ (in
the half of the first Brillouin zone) for different values of the amplitude
$A$ of the stationary periodic solution. Panels (a) and (b) present the
results for the self-focusing ($g=0.3$) and defocusing ($g=-0.7$)
nonlinearity, respectively. Black solid, red short-dashed, blue short-dotted,
gray circled, and green dotted curves correspond to $A=0.1$, $0.5$, $1.0$,
$1.23$, and $1.3$ in (a), and $A=0.1$, $0.35$, $0.95$, $1.03$, and $1.3$ in
(b). The strength of the gain-loss term is $V_{0}=0.05$ in (a) and $V_{0}=0.2$
in (b), respectively, other parameters being $\omega=1$, $\phi=\pi/2$, and
$\mu=-1$.}%
\end{figure}

In contrast with the results displayed in Fig. 1, even if the periodic
solution is unstable against small random-noise perturbations, it may be
effectively stable against perturbations in the form of specific Floquet-Bloch
eigenmodes, due to the existence of the respective stability band, as shown in
Figs. 3(a,c), which we call \textit{band stability}.

Next, Fig. 5 shows the effect of amplitude $A$ of the stationary periodic
solution on its stability. For the self-focusing nonlinearity, as seen in Fig.
5(a), at $A=1.23$ and $A=1.3$ [the gray circled and green dotted curves in
Fig. 5(a)] the periodic solutions are unstable to perturbations with all
values of $k$. At $A=0.1$, $0.5$ and $1.0$ [see black solid, red short-dashed,
and blue short-dotted curves in Fig. 5(a)], stability bands are the intervals
of $k\in\lbrack0,0.49]$, $[0.15,0.46]$, and $[0.29,0.42]$, respectively, whose
widths shrink with the increasing of $A$.

\begin{figure}[ptb]
\centering\vspace{0.0cm} \includegraphics[width=8.5cm]{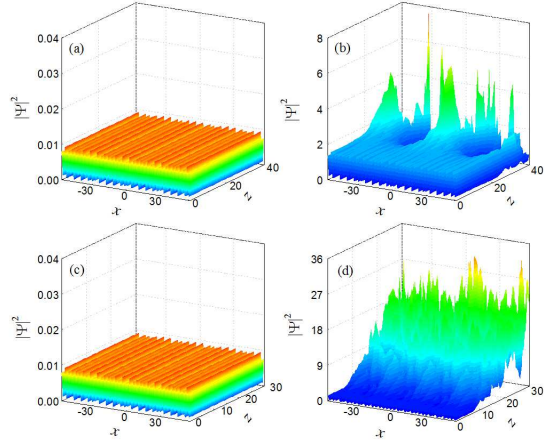}
\vspace{-0.0cm}\caption{(Color online) The simulated evolution of periodic
solutions, perturbed by specially selected eigenmodes (see the text), under
the action of the self-focusing nonlinearity ($g=0.3$) in (a) and (b), and
defocusing nonlinearity ($g=-0.7$) in (c) and (d). The parameters are: (a)
$A=0.1$, $V_{0}=0.05$, $k=0.1$; (b) $A=1.3$, $V_{0}=0.05$, $k=0.1$; (c)
$A=0.1$, $V_{0}=0.2$, $k=0.2$; (d) $A=1.3$, $V_{0}=0.2$, $k=0.2$. The
perturbation amplitude is $\epsilon=0.01$, other parameters being the same as
in Fig. 5.}%
\end{figure}

Similarly, for the defocusing nonlinearity, as seen in Fig. 5(b), at $A=1.03$
and $A=1.3$ the stationary periodic solutions are again unstable to
perturbations with all $k$, see the gray circled and green dotted curves in
Fig. 5(b). At $A=0.1$, $0.35$, and $0.95$, as shown by the black solid, red
short-dashed, and blue short-dashed curves in Fig. 5(b), the stability bands
are the intervals of $k\in\lbrack0.05,0.46]$, $[0.19,0.39]$, and $[0,0.15]$, respectively.

\begin{figure}[ptb]
\centering\vspace{0cm} \includegraphics[width=8.5cm]{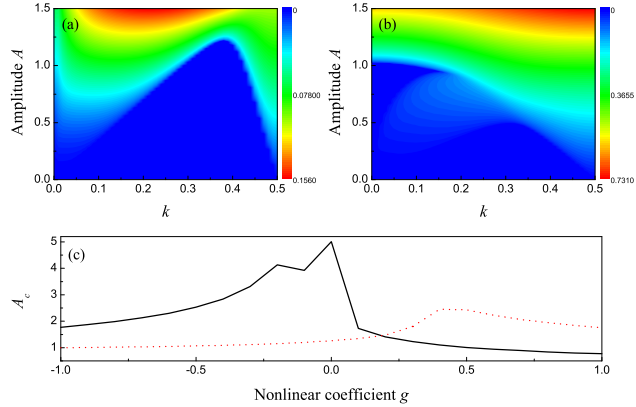} \vspace
{0cm}\caption{(Color online) The dependence of the instability growth rate,
i.e., the largest imaginary part of the eigenvalues, on amplitude $A$ of the
stationary solutions in the half of first Brillouin zone, for (a) $g=0.3$,
$V_{0}=0.05$, and (b) $g=-0.7$, $V_{0}=0.2$. (c) The critical value $A_{c}$,
beyond which the instability occurs at all $k$, versus the nonlinearity
coefficient, $g$. Other parameters are the same as in Fig. 5.}%
\end{figure}

The above results can be confirmed by simulations of Eq. (\ref{Model}) with
inputs in the form of the periodic solution with the addition of small
perturbations corresponding to specific Floquet-Bloch eigenmodes. Figs. 6(a)
and 6(b) show the evolution of the periodic solutions perturbed by the
eigenmodes with $k=0.1$ for $A=0.1$ and $A=1.3$, with the gain/loss
coefficient $V_{0}=0.05$, in the self-focusing regime. As predicted by points
\textquotedblleft a\textquotedblright\ and \textquotedblleft
b\textquotedblright\ in Fig. 5(a), the periodic solution is stable for $A=0.1$
[Fig. 6(a)], and unstable for $A=1.3$ [Fig. 6(b)]. Figs. 6(c) and 6(d) display
the corresponding results in the defocusing regime, where the perturbation
eigenmodes with $k=0.2$ for $A=0.1$ and $A=1.3$ are initially added to the
system with $V_{0}=0.2$. It is seen that the periodic solution is stable at
$A=0.1$, see Fig. 6(c), and it is unstable at $A=1.3$, see Fig. 6(d), in
agreement with the predictions produced by points \textquotedblleft
c\textquotedblright\ and \textquotedblleft d\textquotedblright\ in Fig. 5(b).

\begin{figure}[ptb]
\centering\vspace{-0.5cm} \includegraphics[width=9.0cm]{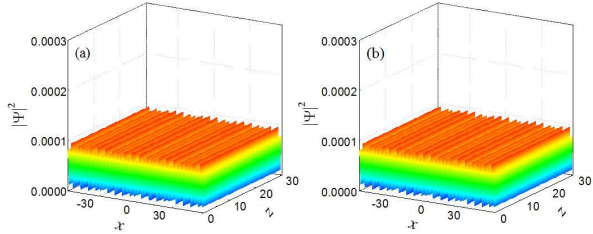}
\vspace{-3.5cm}\caption{(Color online) The evolution of the perturbed periodic
solution under the action of the periodic complex potential $V(x)$ given by
Eqs. (\ref{VR}) and (\ref{VI}) for (a) the self-focusing nonlinearity with
$g=0.3$ and $V_{0}=0.05$; and (b) the defoucsing nonlinearity with $g=-0.7$
and $V_{0}=0.2$. Here, the perturbation amplitude $\varepsilon=0.02$, and the
other parameters are taken as $A=0.01$, $\omega=1$, $\phi=\pi/2$, and $\mu
=-1$.}%
\end{figure}

Further, Fig. 7 shows the dependence of the largest imaginary part of the
eigenvalues on amplitude $A$ of the stationary periodic solutions in the half
of first Brillouin zone for the self-focusing and defocusing\ nonlinearities.
In the former case, Fig. 7(a) shows that the eigenvalues are almost completely
real at all $k$ for small $A$. With the increase of $A$, the stability band
shrinks up to $A=1.23$, beyond which the eigenvalues with the largest
imaginary part are complex at all $k$. Similarly, in the defocusing regime, the periodic solutions are stable at all $k$ for small $A$, and the instability takes place at all values of $k$ for $A>1.03$, as shown in Fig. 7(b). The
dependence of the critical value $A_{c}$, beyond which the instability occurs
at all $k$, on the nonlinearity coefficient $g$ is presented in Fig. 7(c), for
$V_{0}=0.05$ and $0.2$ (the black solid and red dotted curves), respectively.
In particular, in the former case, the dependence $A_{c}(g)$ \ clearly implies
that the instability is essentially stronger for the self-focusing sign of the
nonlinearity ($g>0$), which is quite natural, as the self-focusing pushes the
system towards the breakup of the $\mathcal{PT}$ symmetry by imposing the
modulational instability.

As the result, the periodic solutions are stable at all wavenumber $k$
for small $A$. In this case, we conclude that they will be stable against
small random-noise perturbations, i.e., dynamical stable. A lot of numerical
simulations confirm the conclusion. As the typical example, Fig. 8 exhibits
the stable evolution of the periodic solutions for $A=0.01$, $g=0.3$,
$V_{0}=0.05$ and $A=0.01$, $g=-0.7$, $V_{0}=0.2$, respectively, which the
instability growth rate is $0$ for all $k$, as shown in Figs. 7(a) and 7(b).

\begin{figure}[ptb]
\centering\vspace{0.0cm} \includegraphics[width=8.5cm]{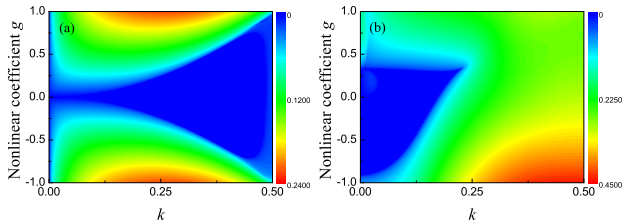}
\vspace{-0.0cm}\caption{(Color online) The dependence of the instability
growth rate, $\max[\operatorname{Im}\lambda(k)]$, on the nonlinearity
coefficient, $g$, and wavenumber, $k$, for (a) $V_{0}=0.01$ and (b)
$V_{0}=0.2$. Here, the parameters are $A=1$, $\omega=1$, $\phi=\pi/2$, and
$\mu=-1$.}%
\end{figure}

Finally, we consider the effect of the nonlinearity coefficient, $g$, on the
stability band of the periodic solution. Fig. 9 displays the dependence of the
instability growth rate on $g$ in the half of the first Brillouin zone for
different values of the gain/loss coefficient, $V_{0}$. In Fig. 9(a), at
$V_{0}=0.01$ the stability band emerges near the edge of the Brillouin zone,
and its size increases with the decrease of $\left\vert g\right\vert $,
reaching a maximum at $g=0$. Note that the respective contour plot is almost
symmetric with respect to $g=0$. However, the symmetry is \ broken at larger
$V_{0}$, as seen in Fig. 9(b), which displays the corresponding results at
$V_{0}=0.2$. They demonstrate, in particular, that the instability occurs at
all $k$ when $g$ belongs to intervals $-1<g<-0.91$ and $0.34<g<1$. At
$-0.91<g<0.34$, the stability band exists, and its size grows with the
increase of $g$.

\section{Conclusion}

We have considered periodic solutions produced by the NLS equation with the
non-Hermitian potential. The model is built with the help of the
inverse-problem approach, which selects the complex periodic potential needed
to support by periodic states with the simple phase-gradient structure. The
setting can be realized in optical nonlinear waveguides, with an appropriate
distribution of the local gain and loss. The analysis was focused on the
linear-stability band structures of the periodic solutions, calculated by
means of the plane-wave-expansion method. The results shows that, even if the
periodic solutions are unstable against small random perturbations, they may
be stable against perturbations of specific Floquet-Bloch eigenmodes, due to
the presence of the stability band in the periodic potential. The approach
elaborated in the present work can be used to analyze stability band
structures of periodic solutions in periodic complex potentials for other
nonlinear systems.

\section*{Acknowledgments}

This research is supported by the National Natural Science Foundation of China
grant 61475198 and 11705108, the Shanxi Scholarship Council of China grant 2015-011.


\section*{References}

\begin{thebibliography}{00}
\bibitem {PRL80_5243}C. M. Bender and S. Boettcher, Real Spectra in
Non-Hermitian Hamiltonians Having $\mathcal{PT}$ Symmetry, Phys.
Rev. Lett. \textbf{80}
(1998) 5243-5246.

\bibitem {JMP43_2814}A. Mostafazadeh, Pseudo-Hermiticity versus $\mathcal{PT}
$-symmetry III: Equivalence of pseudo-Hermiticity and the presence of
antilinear symmetries, J. Math. Phys. \textbf{43} (2002) 3944-3951.

\bibitem {PRL89_270401}C. M. Bender, D. C. Brody, and H. F. Jones, Complex
Extension of Quantum Mechanics, Phys.
Rev. Lett. \textbf{89} (2002) 270401.

\bibitem {Rep.Prog.Phys70_947}C. M. Bender, Making sense of non-Hermitian
Hamiltonians, Rep. Prog. Phys. \textbf{70} (2007) 947-1018.

\bibitem {Guo}A. Guo, G. J. Salamo, D. Duchesne, R. Morandotti, M.
Volatier-Ravat, V. Aimez, G. A. Siviloglou, and D. N. Christodoulides,
Observation of PT-Symmetry Breaking in Complex Optical Potentials,
Phys.
Rev. Lett. \textbf{103} (2009) 093902.

\bibitem {NP6_192}C. E. R\"{u}ter, K. G. Makris, R. El-Ganainy, D. N.
Christodoulides, M. Segev, and D. Kip, Observation of parity-time symmetry in
optics, Nature Phys. \textbf{6} (2010) 192-195; T. Kottos, Broken symmetry
makes light work, Nature Phys. \textbf{6} (2010) 166-167.

\bibitem {Nat488_167}A. Regensburger, C. Bersch, M. -A. Miri, G. Onishchukov,
D. N. Christodoulides, and U. Peschel, Parity-time synthetic photonic
lattices, Nature \textbf{488} (2012) 167-171.

\bibitem {NatMater12_108}L. Feng, Y. -L. Xu, W. S. Fegadolli, M. -H. Lu, J. E.
B. Oliveira, V. R. Almeida, Y. -F. Chen, and A. Scherer, Experimental
demonstration of a unidirectional reflectionless parity-time metamaterial at
optical frequencies, Nature Mater. \textbf{12} (2013) 108-113.

\bibitem {Microres}L. Chang, X. Jiang, S. Hua, C. Yang, J. Wen, and L. Jiang,
G. Li, G. Wang, and Min Xiao, Parity-time symmetry and variable optical
isolation in active-passive-coupled microresonators, Nature Phot. \textbf{8}
(2014) 524-529.

\bibitem {NP10_394}B. Peng, S. Kaya\"{o}zdemir, F. Lei, F. Monifi, M.
Gianfreda, G. L. Long, S. Fan, F. Nori, C. M. Bender, and L. Yang,
Parity-time-symmetric whispering-gallery microcavities Nature Phys.
\textbf{10} (2014) 394-398.

\bibitem {atomic chain}Z. Zhang, Y. Zhang, J. Sheng, L. Yang, M.-A. Miri, D.
N. Christodoulides, B. He, Y. Zhang, and M. Xiao, Observation of Parity-Time
Symmetry in Optically Induced Atomic Lattices, Phys. Rev. Lett. \textbf{117}
(2016) 123601.

\bibitem {Optica}Z. Gao, S. T. M. Fryslie, B. J. Thompson, P. S. Carney, and
K. D. Choquette, Parity-time symmetry in coherently coupled vertical cavity
laser arrays, Optica \textbf{4} (2017) 323-329.

\bibitem {PRL100_103904}K. G. Makris, R. El-Ganainy, D. N. Christodoulides,
and Z. H. Musslimani, Beam Dynamics in PT Symmetric Optical Lattices,
Phys.
Rev. Lett. \textbf{100} (2008) 103904.

\bibitem {PRL103_093902}A. Guo, G. J. Salamo, D. Duchesne, R. Morandotti, M.
Volatier-Ravat, V. Aimez, G. A. Siviloglou, and D. N. Christodoulides,
Observation of PT-Symmetry Breaking in Complex Optical Potentials,
Phys.
Rev. Lett. \textbf{103 }(2009)\textbf{\ }093902.

\bibitem {PRL113_023902}R. Fleury, D. L. Sounas, and A. Al\`{u}, Negative
Refraction and Planar Focusing Based on Parity-Time Symmetric Metasurfaces,
Phys.
Rev. Lett. \textbf{113 }(2014)\textbf{\ }023903.

\bibitem {PRL103_123601}S. Longhi, Bloch Oscillations in Complex Crystals with
PT Symmetry, Phys.
Rev. Lett. \textbf{103} (2009) 123601.

\bibitem {SR5_17760}M. Wimmer, M. Miri, D. Christodoulides, Ulf Peschel,
Observation of Bloch oscillations in complex PT-symmetric photonic lattices,
Scientific Reports \textbf{5} (2015) 17760.

\bibitem {PRA82_043803}H. Ramezani, T. Kottos, R. El-Ganainy, and D. N.
Christodoulides, Unidirectional nonlinear PT-symmetric optical structures,
Phys.
Rev. A \textbf{82} (2010) 043803.

\bibitem {PRL106_213901}Z. Lin, H. Ramezani, T. Eichelkraut, T. Kottos, H.
Cao, and D. N. Christodoulides, Unidirectional Invisibility Induced by
PT-Symmetric Periodic Structures, Phys.
Rev. Lett. \textbf{106} (2011) 213901.

\bibitem {PRL110_234101}N. Bender, S. Factor, J. D. Bodyfelt, H. Ramezani, D.
N. Christodoulides, F. M. Ellis, and T. Kottos, Observation of Asymmetric
Transport in Structures with Active Nonlinearities, Phys.
Rev. Lett. \textbf{110} (2013) 234101.

\bibitem {PRL112_143903}Y. Sun, W. Tan, H. Q. Li, J. Li, and H. Chen,
Experimental Demonstration of a Coherent Perfect Absorber with PT Phase
Transition, Phys.
Rev. Lett. \textbf{112} (2014) 143903.

\bibitem {Science346_975}H. Hodaei, M. -A. Miri, M. Heinrich, D. N.
Christodoulides, and M. Khajavikhan, Parity-time-symmetric microring lasers,
Science \textbf{346} (2014) 975-978.

\bibitem {Science346_972}L. Feng, Z. J. Wong, R. -M. Ma, Y. Wang, and X.
Zhang, Single-mode laser by parity-time symmetry breaking, Science
\textbf{346} (2014) 972-975.

\bibitem {PRL117_224302}A. V. Poshakinskiy, A. N. Poddubny, and A. Fainstein,
Multiple Quantum Wells for $\mathcal{PT}$-Symmetric Phononic Crystals,
Phys.
Rev. Lett. \textbf{117} (2016) 224302.

\bibitem {NatPhys}L. Feng, R. El-Ganainy, and L. Ge, Non-Hermitian photonics
based on parity-time symmetry, Nature Phot. \textbf{11} (2017) 752-762.

\bibitem {18}Z. H. Musslimani, K. G. Makris, R. El-Ganainy, and D. N.
Christodoulides, Optical Solitons in PT Periodic Potentials, Phys.
Rev. Lett. \textbf{100}
(2008) 030402.

\bibitem {19}F. Kh. Abdullaev, Y. V. Kartashov, V. V. Konotop, and D. A.
Zezyulin, Solitons in PT -symmetric nonlinear lattices, Phys.
Rev. A \textbf{83} (2011) 041805(R).

\bibitem {Mirosh}A. E. Miroshnichenko, B. A. Malomed, and Y. S. Kivshar,
Nonlinearly $\mathcal{PT}$-Symmetric systems: Spontaneous symmetry breaking
and transmission resonances, Phys. Rev. A \textbf{84} (2011) 012123.

\bibitem {22}S. Hu, X. Ma, D. Lu, Z. Yang, Y. Zheng, and W. Hu, Solitons
supported by complex PT -symmetric Gaussian potentials, Phys.
Rev. A \textbf{84} (2011) 043818.

\bibitem {20}X. Zhu, H. Wang, L. X. Zheng, H. G. Li, and Y. J. He, Gap
solitons in parity-time complex periodic optical lattices with the real part
of superlattices, Opt. Lett. \textbf{36} (2011) 2680-2682.

\bibitem {21}H. G. Li, Z. W. Shi, X. J. Jiang, and X. Zhu, Gray solitons in
parity-time symmetric potentials, Opt. Lett. \textbf{36} (2011) 3290-3292.

\bibitem {32}L. Feng, M. Ayache, J. Huang, Y. -L. Xu, M. -H. Lu, Y. -F. Chen,
Y. Fainman, and A. Scherer, Nonreciprocal Light Propagation in a Silicon
Photonic Circuit, Science \textbf{333} (2011) 729-733.

\bibitem {23}S. Nixon, L. Ge, and J. Yang, Stability analysis for solitons in
PT -symmetric optical lattices, Phys.
Rev. A \textbf{85} (2012) 023822.

\bibitem {24}D. A. Zezyulin and V. V. Konotop, Nonlinear modes in the harmonic
$\mathcal{PT}$-symmetric potential, Phys.
Rev. A \textbf{85} (2012) 043840.

\bibitem {OL37_4543}C. Li, C. Huang, H. Liu, and L. Dong, Multipeaked gap
solitons in $\mathcal{PT}$-symmetric optical lattices, Opt. Lett. \textbf{37} (2012) 4543-4545.

\bibitem {25}V. Achilleos, P. G. Kevrekidis, D. J. Frantzeskakis, and R.
Carretero-Gonz\'{a}lez, Dark solitons and vortices in $\mathcal{PT}$-symmetric
nonlinear media: From spontaneous symmetry breaking to nonlinear
$\mathcal{PT}$ phase transitions, Phys.
Rev. A \textbf{86} (2012) 013808.

\bibitem {26}M. -A. Miri, A. B. Aceves, T. Kottos, V. Kovanis, and D. N.
Christodoulides, Bragg solitons in nonlinear $\mathcal{PT}$-symmetric periodic
potentials, Phys.
Rev. A \textbf{86} (2012) 033801.

\bibitem {PRA87_013812}Y. He and D. Mihalache, Lattice solitons in optical
media described by the complex Ginzburg-Landau model with $\mathcal{PT}%
$-symmetric periodic potentials,Phys.
Rev. A \textbf{87 }(2013) 013812.

\bibitem {27}B. Midya and R. Roychoudhury, Nonlinear localized modes in
$\mathcal{PT}$-symmetric Rosen-Morse potential wells, Phys.
Rev. A \textbf{87} (2013) 045803.

\bibitem {29}N. Lazarides and G. P. Tsironis, Gain-Driven Discrete Breathers
in $\mathcal{PT}$-Symmetric Nonlinear Metamaterials, Phys.
Rev. Lett. \textbf{110} (2013) 053901.

\bibitem {30}G. Castaldi, S. Savoia, V. Galdi, A. Al\`{u}, and N. Engheta,
$\mathcal{PT}$ Metamaterials via Complex-Coordinate Transformation Optics,
Phys.
Rev. Lett. \textbf{110} (2013) 173901.

\bibitem {28}C. P. Jisha, L. Devassy, A. Alberucci, and V. C. Kuriakose,
Influence of the imaginary component of the photonic potential on the
properties of solitons in $\mathcal{PT}$-symmetric systems, Phys.
Rev. Lett. \textbf{90}
(2014) 043855.

\bibitem {31}A. Lupu, H. Benisty, and A. Degiron, Switching using PT symmetry
in plasmonic systems: positive role of the losses, Opt. Exp. \textbf{21}
(2013) 21651-21668.

\bibitem {Raymond}Z. Chen, J. Liu, S. Fu, Y. Li, and B. A. Malomed, Discrete
solitons and vortices on two-dimensional lattices of $\mathcal{PT}$-Symmetric
couplers, Opt. Exp. \textbf{22} (2014) 29679-29692.

\bibitem {OC313_139}J. Xie, Z. Su, W. Chen, G. Chen, J. Lv, D. Mihalache, and
Y. He, Defect solitons in two-dimensional photonic lattices with
parity-timesymmetry, Opt. Commun. \textbf{313\ }(2014) 139-145.

\bibitem {OC315_303}C. Dai and Y. Wang, Nonautonomous solitons in parity-time
symmetric potentials, Opt. Commun. \textbf{315\ }(2014)303-309.

\bibitem {PRA90_034833}C. Huang, F. Ye, and X. Chen, Mode pairs in
$\mathcal{PT}$-symmetric multimode waveguides, Phys.
Rev. A \textbf{90} (2014) 043833.

\bibitem {OC335_146}H. Wang, S. Shi, X. Ren, X. Zhu, B. A. Malomed, D.
Mihalache, and Y. He, Two-dimensional solitons in triangular photonic lattices
with parity-timesymmetry, Opt. Commun. \textbf{335} (2015) 146-152.

\bibitem {RMP}V. V. Konotop, J. Yang, and D. A. Zezyulin, Nonlinear waves in
$\mathcal{PT}$-symmetric systems, Rev. Mod. Phys. \textbf{88} (2016) 035002.

\bibitem {Suchkov}S. V. Suchkov, A. A. Sukhorukov, J. Huang, S. V. Dmitriev,
C. Lee, and Y. S. Kivshar, Nonlinear switching and solitons in PT-symmetric
photonic systems, Laser Photonics Rev. \textbf{10} (2016) 177-213.

\bibitem {unbreakable}Y. V. Kartashov, B. A. Malomed, and L. Torner,
Unbreakable PT symmetry of solitons supported by inhomogeneous defocusing
nonlinearity, Opt. Lett. \textbf{39 }(2014) 5641-5644.

\bibitem {PRA86_013612}H. Cartarius and G. Wunner, Model of a $\mathcal{PT}%
$-symmetric Bose-Einstein condensate in a $\delta$-function double-well
potential, Phys.
Rev. A \textbf{86} (2012) 013612.

\bibitem {PRA90_042123}F. Single, H. Cartarius, G. Wunner, and J. Main,
Coupling approach for the realization of a PT -symmetric potential for a
Bose-Einstein condensate in a double well, Phys.
Rev. A \textbf{90} (2014) 042123.

\bibitem {PRA93_033617}D. Dast, D Haag, H. Cartarius, and G. Wunner, Purity
oscillations in Bose-Einstein condensates with balanced gain and loss,
Phys.
Rev. A \textbf{93} (2016) 033617.

\bibitem {PRA91_043629}Y. Zhang, Y. Xu, and T. Busch, Gap solitons in
spin-orbit-coupled Bose-Einstein condensates in optical lattices,
Phys.
Rev. A \textbf{91} (2015) 043629.

\bibitem {OL38_4033}C. Hang, D. A. Zezyulin, V. V. Konotop, and G. Huang,
Tunable nonlinear parity-time-symmetric defect modes with an atomic cell,
Opt. Lett. \textbf{38} (2013) 4033-4036.

\bibitem {OL39_5387}C. Hang, D. A. Zezyulin, G. Huang, V. V. Konotop, and B.
A. Malomed, Tunable nonlinear double-core PT-symmetric waveguides,
Opt. Lett. \textbf{39}, 5387-5390 (2014).

\bibitem {OE21_32053}H. Li, J. Dou, and G. Huang, PT symmetry via
electromagnetically induced transparency, Optics Express \textbf{21} (2013) 32053-32062.

\bibitem {NJP18_065001}M. Miri and Andrea Al\`{u}, Nonlinearity-induced
PT-symmetry without material gain, New J. Phys. \textbf{18} (2016) 065001.

\bibitem {SH}H. Sakaguchi and B. A. Malomed, Gap solitons in Ginzburg-Landau
media, Phys.
Rev. E \textbf{77} (2008) 056606.

\bibitem {OL41_2747}S. Nixon and J. Yang, Nonlinear light behaviors near phase
transition in non-parity-time-symmetric complex waveguides, Opt. Lett.
\textbf{41} (2016) 2747-2750.

\bibitem {PhysicaD331_48}S. Nixon and J. Yang, Nonlinear wave dynamics near
phase transition in PT-symmetric localized potentials, Physica D \textbf{331}
(2016) 48-57.

\bibitem {PLA380_3803}J. Yang and S. Nixon, Stability of soliton families in
nonlinear Schr\"{o}inger equations with non-parity-time-symmetric complex
potentials, Phys. Lett. A \textbf{380 }(2016) 3803-3809.

\bibitem {PRL105_053901}Y. D. Chong, L. Ge, H. Cao, and A. D. Stone, Coherent
Perfect Absorbers: Time-Reversed Lasers, Phys.
Rev. Lett. \textbf{105} (2010) 053901.

\bibitem {Science331_889}W. Wan, Y. Chong, L. Ge, H. Noh, A. D. Stone, and H.
Cao, Time-Reversed Lasing and Interferometric Control of Absorption, Science
\textbf{331} (2011) 889-892.

\bibitem {PRL108_173901}M. Liertzer, L. Ge, A. Cerjan, A. D. Stone, H. E.
T\"{u}reci, and S. Rotter, Pump-Induced Exceptional Points in Lasers,
Phys.
Rev. Lett. \textbf{108} (2012) 17390.

\bibitem {NC5_4034}M. Brandstetter, M. Liertzer, C. Deutsch, P. Klang, J.
Sch\"{o}berl, H. E. T\"{u}reci, G. Strasser, K. Unterrainer, and S. Rotter,
Reversing the pump dependence of a laser at an exceptional point, Nat. Commun
\textbf{5} (2014) 4034.

\bibitem {science346_328}B. Peng, S. K. \"{O}zdemir, S. Rotter, H. Yilmaz, M.
Liertzer, F. Monifi, C. M. Bender, F. Nori, L. Yang, Loss-induced suppression
and revival of lasing, Science \textbf{346} (2014) 328-332.

\bibitem {PRA95_023833}C. Hang, G. Gabadadze, and G. Huang, Realization of
non-PT-symmetric optical potentials with all-real spectra in a coherent atomic
system, Phys.
Rev. A \textbf{95} (2017) 023833.

\bibitem {PRE83_036608}K. Li and P. G. Kevrekidis, $\mathcal{PT}$-symmetric
oligomers: Analytical solutions, linear stability, and nonlinear dynamics,
Phys.
Rev. E \textbf{83} (2011) 066608.

\bibitem {OL36_4323}R. Driben and B. A. Malomed, Stability of solitons in
parity-time-symmetric couplers, Opt. Lett. \textbf{36} (2011) 4323-4325.

\bibitem {J.Opt15_064010}Y. V. Bludov, R. Driben, V. V. Konotop, and B. A.
Malomed, Instabilities, solitons and rogue waves in $\mathcal{PT}$-coupled
nonlinear waveguides, J. Opt. \textbf{15 }(2013) 064010.

\bibitem {oe22_19774}X. Ren, H. Wang, H. Wang, and Y. He, Stability of
in-phase quadruple and vortex solitons in the parity-time-symmetric periodic
potentials, Opt. Express \textbf{22 }(2014)\textbf{\ }19774-19782.

\bibitem {PRE91_023203}L. Ge, M. Shen, T. Zang, C. Ma, and L. Dai, Stability
of optical solitons in parity-time-symmetric optical lattices with competing
cubic and quintic nonlinearities, Phys.
Rev. E \textbf{91} (2015) 023203.

\bibitem {PhysicaD313_26}J. T. Cole and Z. H. Musslimani, Spectral transverse
instabilities and soliton dynamics in the higher-order multidimensional
nonlinear Schr\"{o}inger equation, Physica D \textbf{313} (2015) 26-36.

\bibitem {PRE92_022913}Z. Yan, Z. Wen, and C. Hang, Spatial solitons and
stability in self-focusing and defocusing Kerr nonlinear media with
generalized parity-time-symmetric Scarff-II potentials, Phys.
Rev. E \textbf{92} (2015) 022913.

\bibitem {NC6_7257}K. G. Makris, Z. H. Musslimani, D. N. Christodoulides and
S. Rotter, Constant-intensity waves and their modulation instability in
non-Hermitian potentials, Nat. Comm. \textbf{6} (2015) 7257.

\bibitem {EPJD71_140}B. Liu, L. Li, and B. A. Malomed, Effects of the
third-order dispersion on continuous waves in complex potentials, Eur. Phys.
J. D \textbf{71} (2017) 140.

\bibitem {semicond}J. R. Marciante and G. P. Agrawal, Nonlinear mechanisms of
filamentation in broad-area semiconductor lasers, IEEE J. Quant. Electr.
\textbf{32} (1996) 590-596.

\bibitem {inverse1}L. D. Carr, C. W. Clark, and W. P. Reinhardt, Stationary
solutions of the one-dimensional nonlinear Schr\"{o}dinger equation. I. Case
of repulsive nonlinearity, Phys. Rev. A \textbf{62} (2000) 063610.

\bibitem {inverse2}L. D. Carr, C. W. Clark, and W. P. Reinhardt, Stationary
solutions of the one-dimensional nonlinear Schr\"{o}dinger equation. II. Case
of attractive nonlinearity, Phys. Rev. A \textbf{62} (2000) 063611.

\bibitem {inverse3}J. Belmonte-Beitia, V. M. P\'{e}rez-Garc\'{\i}a, V.
Vekslerchik, and V. V. Konotop, Localized Nonlinear Waves in Systems with
Time- and Space-Modulated Nonlinearities, Phys. Rev. Lett. \textbf{100} (2008) 164102.

\bibitem {inverse4}B. A. Malomed and Yu. A. Stepanyants, The inverse problem
for the Gross-Pitaevskii equation, Chaos \textbf{20} (2010) 013130.

\bibitem {inverse5}E. Ding, H. N. Chan, K. W. Chow, K. Nakkeeran and B. A.
Malomed, Exact states in waveguides with periodically modulated nonlinearity,
Exact states in waveguides with periodically modulated nonlinearity, EPL
\textbf{119} (2017) 54002.

\bibitem {VVK}D. A. Zezyulin, I. V. Barashenkov, and V. V. Konotop, Stationary
through-flows in a Bose-Einstein condensate with a PT-symmetric impurity,
Phys.
Rev. A \textbf{94} (2016) 063649.
\end{thebibliography}
\end{document}